# Secure FSM-based Arithmetic Codes

Hashem Moradmand Ziyabar[1], Mahnaz Sinaie[2], Ali Payandeh[3], Vahid Tabataba Vakili[4]

[1.] Sazgan Ertebat Co. , [2.] Tarbiat Modares University , [3.] MalekAshtar University of Technology,

[4.] Iran University of Science and Technology

Corresponding author: Hashem Moradmand Ziyabar, h.moradmand@gmail.com

**Abstract**— recently, arithmetic coding has attracted the attention of many scholars because of its high compression capability. Accordingly, in this paper a method which adds secrecy to this well-known source code is proposed. Finite state arithmetic code (FSAC) is used as source code to add security. Its finite state machine (FSM) characteristic is exploited to insert some random jumps during source coding process. In addition, a Huffman code is designed for each state to make decoding possible even in jumps. Being Prefix free, Huffman codes are useful in tracking correct states for an authorized user when s/he decodes with correct symmetric pseudo random key. The robustness of our proposed scheme is further reinforced by adding another extra uncertainty by swapping outputs of Huffman codes in each state. Several test images are used for inspecting the validity of the proposed Huffman Finite State Arithmetic Coding (HFSAC). The results of several experimental, key space analyses, statistical analysis, key sensitivity and plaintext sensitivity tests show that HFSAC with a little effect on compression efficiency for image cryptosystem provides an efficient and secure way for real-time image encryption and transmission.

## 1. Introduction

The efficiency and security requirements of information transmission have led to conduct a substantial amount of research on data compression and encryption. In order to improve the performance and the flexibility of multimedia applications, it is worth performing compression and encryption in a single process [1]-[6]. Recently, arithmetic coding has attracted the attention of many scholars due to the high compression efficiency in the applications [5] such as JPEG2000 and H.264.

Grangetto et.al [1], [2] introduced a randomized arithmetic coding scheme which achieves security by random changing of the symbol intervals. Kim et.al [4], [5] inserted encryption by splitting coding interval according to a random key. Some secure arithmetic codes were proposed by Bose et.al and Chen et. al. [6], [7] who used chaos theory to make randomized arithmetic encoder model. Howard et.al [8],[9] introduced finite state arithmetic coding and showed that this implementation didn't affect compression efficiency. While the previous research has used integer arithmetic coding, the present study benefited from finite state integer arithmetic codes (FSAC) with a symmetric pseudo random key to add security to this source code. We used a secure random key to jump to some states. Moreover, we defined Huffman codes for each state to prevent fault decoding. The numerical results showed that the proposed method satisfied the security terms.

This paper is organized as follows. In section 2, the studies conducted on finite state integer arithmetic codes are briefly reviewed. The secure arithmetic coding proposed in this research is presented in Section 3. Subsequently, sections 4 and 5 offers the implications and performance analysis of the proposed approach and section 6 offers conclusion and some suggestions for interested readers for exploring the related issues to this study.

# 2. Brief review of FSAC

## 2.1. FSAC algorithm

Arithmetic coding is the process of recursively selecting subintervals according to the probability of the next coming symbols [7]. Pure arithmetic codes need infinite precision and have no output until the end of the block is being encoded. To determine the finite precision, integer arithmetic codes with incremental outputs were introduced [8], [10]. FSAC is a revised version of integer arithmetic codes having a finite number of states [11]. The process of FSAC with precision of N (an integer number to indicate the length of the initial interval) and Fmax (maximum value of follow count) is described in Table I.

Table I. Finite state Integer arithmetic encoding

```
Initialize current interval [low,high] = [0,N) and follow = 0
for each input symbol:
    Divide current interval into subintervals proportional to
    the probability of symbols.
    Current Interval = subinterval of coming symbol.
    Repeatedly do:
        If new subinterval ∉ [0,N/2) or [N/4,3N/4), or
        [N/2,N), exit and return.
        Else if {(follow >= Fmax) AND (high∈[N/2,3N/2)
        AND low ∈ is in [N/4,N/2)}
            switch(next symbol)
                case 0: low=N/2
                case 1 :high=N/2-1
        If current interval ∈ [0,N/2),
            output 0 and any following 1's left
            behind, double the size of the
            subinterval by linearly expanding
            [0,N/2) to [0,N).
        If current interval ∈[N/2,N),
            output 1 and any following 0's left
            behind,double the size of the
            subinterval by linearly expanding
            [N/2,N) to [0,N).
        If current interval∈[N/4,3N/4),
            increment the follow by one; double
            the size of the subinterval by linearly
            expanding [N/4,3N/4) to [0,N).
Output enough bits to distinguish the final current
```

Table II. Number of states with different N

| Encoder Number # | i | ii | iii | iv |
|---|---|---|---|---|
| N | 3 | 5 | 7 | 7 |
| Fmax | 1 | 1 | 1 | 10 |
| Reduced States # | 1 | 12 | 172 | 203 |
| Reduced Transitions # | 4 | 57 | 732 | 2825 |
| Interval Division # | [2 6] | [6 26] | [26 102] | [26 102] |

## 2.2. State diagram of FSAC

Being described with finite number of states, FSAC encoder can be shown as a state diagram. Such a state diagram for N=3, Fmax=1 and the probability P(0) =3/8 for a binary source is shown in Fig.1. The numbers that are written on the edges are the inputs/outputs of each transition. This state diagram has 4 states and 10 transitions between the states. Three of transitions have no outputs which are called mute transitions. If the code is designed for a relatively large value of N and/or Fmax, the number of states increases rapidly; to make this point more clear, some binary state machines with different precisions N, are summarized in Table II.

The next step to more simplifying the arithmetic encoder is to reduce the state diagram by removing the mute transitions [11]. In brief, the process of reduction is to combine mute transitions in other transitions and finally to remove extra states. For example transition 1/- in Fig.1.a from state 0 to state 2 can be combined in two transitions out of state 2, making two self-returning transitions of state 0 in Fig.1.b, i.e. 10/011 and 11/1. The final reduced state diagram of the previous example is shown in Fig.1.b.

As expected, the *inputs* of encoder are prefix-free. For example in state zero the inputs are {11,10,0} where, none of them is prefix of the others. Clearly, being prefix free is a necessity for decodable codes. On the other hand, not only the *outputs* are not prefix-free [10] but also they are related to each other. This is because of block coding behavior of arithmetic codes which boosts their compression capability.

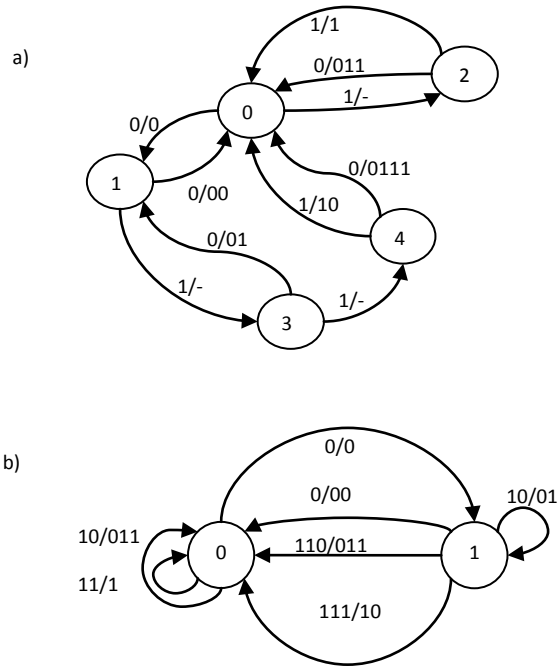

Figure1. Full state transition of encoder (a) and its reduced diagram (b)

## 3. Proposed Method

The present paper aims to provide a secure arithmetic coding system. Our method is based on FSM of the encoder. We used the random output sequence of a Pseudo Random Number Generator (PRNG) as a symmetric key to select the next state during the en/decoding process. In addition, a Huffman tree was defined for each state according to each state's arithmetic code output length. This way, deciphering and tracking the states is only possible for receivers having the correct key. The details will be discussed more in this section. While all of the methods described here can be applied for coding of source alphabets with any size, the authors address the case of binary systems here to simplify the discussion and illustrations.

### 3.1. Inserting Pseudo Random Jumps

The method proposed in this study is based on jumping among states by using a symmetric key which is called *jump key* thereafter. In this method, at some specific time instances, enforced by jump key, the subsequent states would not be chosen according to ordinary state flow of FSM. An example of these kinds of jump for 8 steps of encoding for Encoder iii in Table II is shown in Table III.

The jump key is composed of two keys; the first key (K1) describes whether to jump or not, thus it is a binary sequence in which ones indicate jumping and zeros indicate no change in encoding process. The second key (K2) determines the following state; it is a pseudo-random sequence with elements from set of all possible states. Finally the jump key can be considered as product of these two keys.

To strengthen the encryption against attackers, the first element of K1 is always defined as *one*; i.e. we always have a jump in start of encoding which breaks the common behavior of starting from state zero. Therefore the attacker cannot track the encoding process by starting from state zero.

In jump points, the next state is changed according to jump key. Clearly, when no jumping occurs, next states are determined with next coming symbols as usual and shown by X's in the table.

Jumping by a random key introduces a problem. The compression capability of arithmetic code arises from coding blocks of symbols instead of individual ones. So there are relations among the outputs of different input symbols in an arithmetically coded sequence and necessarily there is not an output for each input symbol. In other words the FSM of encoder has memory and breaking the output sequence (jumping) would lead to losing the decoding track. To solve this problem, we tried to make each state independent of next one. Thus we defined a Huffman code for each state on the basis of the output length of the arithmetic code of that particular state to attain another set of outputs. We assumed the arithmetic code output length, for each symbol is heuristically a measure of its probability. Consequently, we take each state individually as a simple source and the length of the output codes as a measure of probability of the symbols as Eq.1, to design a proper Huffman code.

$$p_{heuristic} \triangleq 2^{-code\ length} \quad (1)$$

State zero of encoder iii in Table II has outputs {1, 011, and 0}. Hence an approximate set of probabilities is {1/2, 1/8, 1/2} which is normalized to {0.44, 0.11, 0.44} and the Huffman code for this set of probabilities would be {1, 01, 11}. Table IV shows Huffman output for states one and two. Huffman code is not a recursive code and each output is directly related to its corresponding input, so jumping among states does not cause any problem, however some portion of compression efficiency is lost. We call this new code as Huffman- output FSAC (HFAC).

Table III. Jump based on jump key

| Instant | 1 | 2 | 3 | 4 | 5 | 6 | 7 | 8 |
|---|---|---|---|---|---|---|---|---|
| K1 | 1 | 0 | 0 | 0 | 1 | 0 | 1 | 0 |
| K2 | 45 | 14 | 2 | 85 | 251 | 200 | 5 | 155 |
| Jump key (K1×K2) | 45 | 0 | 0 | 0 | 251 | 0 | 5 | 0 |
| Ordinary State Sequence | 3 | 120 | 23 | 46 | 54 | 100 | 102 | 212 |
| Encryption State Sequence | 45 | X | X | X | 251 | X | 5 | X |

Table IV. Designing Huffman code for each state

| Transition No# | Current State | Next State | Input | Output | Huffman Output |
|---|---|---|---|---|---|
| 1 | State 0 | 0 | 11 | 1 | 1 |
| 2 | State 0 | 0 | 10 | 011 | 01 |
| 3 | State 0 | 1 | 0 | 0 | 00 |
| 4 | State 1 | 1 | 10 | 01 | 01 |
| 5 | State 1 | 0 | 0 | 00 | 10 |
| 6 | State 1 | 0 | 110 | 0111 | 11 |
| 7 | State 1 | 0 | 111 | 10 | 00 |

Table V. Encoder table of N=4, p (0) = 0.2, Fmax=1

| Current State | Input | Huffman Output | Next State |
|---|---|---|---|
| 0 | 0 | 000 | 1 |
| 0 | 10 | 0011 | 0 |
| 0 | 110 | 01 | 2 |
| 0 | 1110 | 0010 | 0 |
| 0 | 1111 | 1 | 0 |
| 1 | 0 | 100 | 0 |
| 1 | 10 | 011 | 0 |
| 1 | 110 | 010 | 1 |
| 1 | 1110 | 1011 | 0 |
| 1 | 11110 | 11 | 2 |
| 1 | 111110 | 1010 | 0 |
| 1 | 111111 | 00 | 0 |
| 2 | 0 | 000 | 1 |
| 2 | 10 | 0011 | 0 |
| 2 | 110 | 01 | 2 |
| 2 | 1110 | 0010 | 0 |
| 2 | 1111 | 1 | 0 |

As stated before because Huffman code is prefix free it is possible to track decoding even in jump points. Consider we are in state zero of encoder of Table IV and the encoded string is as 01101101…; clearly if we take first bit 0 of string, it will not be any of Huffman outputs of this state therefore we take the next bit, 1, so we have 01 which clearly implies that corresponding input is 10. Now let's try this stream with ordinary arithmetic output of the encoder. We can take one bit, 0, and go to state one or yet we can take three bits, 011, and stay in the same state. This ambiguity arises from non prefix free property of ordinary arithmetic code which explains why we derive Huffman output for each state.

### 3.2. Improving Statistical Properties

To add more security to the encryption procedure, another key was added to the system which is called *swap key* (K3) thereafter. This second key will act on Huffman outputs in each state. The effect of swap key is to remove the iterative patterns in the output which are used by chosen plaintext attacks. Swap key selects a bit position in each Huffman output sequence and reverses all the next bits. To generate the swap key, we used another PRNG, which generates numbers from zero to the maximum length of Huffman output sequences for that state. Consider an encoder which has these parameters: N=4, Fmax=1 and P (0) = 0.2. Outputs of three states for this encoder by use of HFAC are shown in Table V. State zero has the outputs with lengths: 3, 4, 2, 4 and 1. As a result, the maximum output length is 4 and, consequently, the swap key will be a random number of the set {0, 1, 2, 3, and 4}.

Assume the swap key is 1 which implies a swap in bit position 1, the Huffman table changes in the lower part of the table as shown bold in Table VI. And the related trees are shown in Fig.2 and Fig.3. Swapping the tree will lead to have completely different code words. For example if the code word is 0010, it will turn to 0101 after swapping as displayed in Table VI.

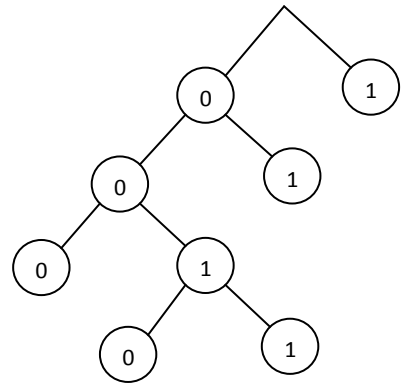

Figure2. Huffman tree of state zero of Table V.

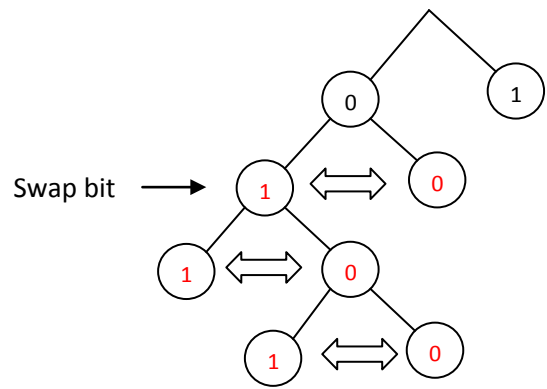

Figure.3. Swapped Huffman tree of Table V.

Table VI. The effect of swapping output

| Bit0 | Bit1 | Bit2 | Bit3 |
|------|------|------|------|
| 0 | 0 | 0 | |
| 0 | 0 | 1 | 1 |
| 0 | 1 | | |
| 0 | 0 | 1 | 0 |
| 1 | | | |
| | ⇧ Swap bit position | | |
| 0 | **1** | 1 | |
| 0 | **1** | 0 | 0 |
| 0 | **0** | | |
| 0 | **1** | 0 | 1 |
| 1 | | | |

A kind of chosen plaintext attack is to choose a sequence which traps the encoder in some limited number of states. Attacker may choose all zeros input sequence. For example consider encoder of Table VII. If the encoder starts from each of three possible states, the state sequences will be {0,1,0,1,…}, {1,0,1,0,…}, and {2,1,0,1,0,…} respectively, i.e. after few bits of input encoder state will alternate between two fixed states, state zero and one. Consequently, any other pattern rather than alternative zeros and ones provided by the key would be a clue for attacker. However, after adding the swap key, if the attacker tries to trap the encryption system in a limited number of states through chosen plaintext, the output manner will not let him to track jump key and to break the system. An observation of Table VII shows these three scenarios in shaded bits. In other words the swapped output sequence will definitely generate a different sequence.

Table VII. Response of encoder in Table V to all zeros input sequence

| State | 0 | 1 | 2 |
|---|---|---|---|
| Input | 0 ,0 ,0 ,… | 0 ,0 ,0 ,… | 0 ,0 ,0 ,… |
| State Sequence | 0 ,1 ,0 ,… | 1 ,0 ,1 ,… | 2 ,1 ,0 ,… |
| Output | 000,100,000,… | 100,000,100,… | 000,100,000,… |
| Swapped Output | 011,101,011,… | 111,111,011,… | 001,101,011,… |

# 4. Simulation results

In this section, we initially discuss the compression efficiency of proposed method; next we elaborate on some simulations to examine the security of proposed method. The pc set we used for simulations is a personal computer with 2G of RAM and Intel Centrino Core 2 Duo 2.2G CPU. Our proposed scheme has been implemented with Matlab software.

## 4.1. Compression efficiency

Indicated in section 3.1, losing some of compression efficiency, is a drawback of HFAC comparing with FASC. We examined HFAC for different precisions and probabilities, the results indicated that the HFAC decreases compression at most (i.e. unequal probabilities) about 10% in comparison to FSAC and integer arithmetic codes. For instance, the outcomes of five sample encoders are shown in Table VIII.

Table VIII. Percent of compression rates

| Encoder | Number of States | AC % | FSAC % | HFAC % |
|---|---|---|---|---|
| P(0)=0.2, N=8, Fmax=1 | 465 | 28.1 | 19.4 | 18.5 |
| P(0)=0.1, N=8, Fmax=3 | 330 | 54.3 | 53.4 | 43.2 |
| P(0)=0.2, N=8, Fmax=3 | 548 | 27.8 | 25.8 | 19.3 |
| P(0)=0.3, N=8, Fmax=3 | 1028 | 11.4 | 7.4 | 1.8 |
| P(0)=0.5, N=8, Fmax=3 | 1 | - 0.6 | - 0.8 | - 0.8 |

## 4.2. Security analysis

A good encryption procedure should be robust enough against all kinds of cryptanalytic, statistical, and brute-force attacks. This includes statistical analysis, key space analysis, and sensitivity analysis [13] which finally verifies high security of the method/scheme against most attackers.

### 4.2.1. Key space

There are two symmetric keys in the proposed encryption system: jump key and swap key. As mentioned earlier, jump key can be treated as pair-wise product of two keys, K1 shows jump points and

K2 shows which states to jump. In simulations K1 is set 50% of times equal to 1 so 50% of times jump occurs. Therefore, if n is the length of plaintext then the K1 space has $\binom{n}{n/2}$ components. In which $\binom{n}{n/2}$ is combinations of $\frac{n}{2}$ out of n. If S is the number of states in the encoder, there will be different possible jump keys as: $S^{\binom{n}{n/2}} = S^{\overline{[(\frac{n}{2})!]^2}}$ Stirling's approximation- $n! \approx n^n e^{-n} \sqrt{2\pi n}$ when n is large enough- is used to more simplifications thus we achieve size of jump key set as Eq.3:

$$\|K1 \times K2\| = S^{\frac{n!}{[(\frac{n}{2})!]^2}} \approx S^{\frac{n^n e^{-n}\sqrt{2\pi n}}{((n/2)^{n/2} e^{-n/2}\sqrt{2\pi n/2})^2}} = S^{\frac{2^n \sqrt{2}}{\sqrt{\pi n}}} \cong S^{0.8 \frac{2^n}{\sqrt{n}}} \quad (3)$$

It is clear that key space grows unboundedly with n, even for typical S values between 20 and 300. So brute-force attack on jump key would be impossible to great extent. If we consider that first state in K1 is always enforced to be one, then K1 space has $\binom{n-1}{\frac{n}{2}-1}$ elements and $\|K1 \times K2\| \cong S^{0.4 \frac{2^n}{\sqrt{n}}}$, therefore abovementioned claim of unbounded key space applies again.

### 4.2.2. Randomness of encrypted stream

One condition for an encrypted sequence to be secure is to have good statistical properties such as entropy. The famous entropy formula for/of a message source, that is, H(S), is displayed below in Eq.4.

$$H(S) = \sum_{s_i} P(s_i) \log_2 \frac{1}{P(s_i)} \text{ bits, } (4)$$

where $P(s_i)$ represents the probability of symbol $s_i$. The entropy is expressed in bits. If the source emits output from set of symbols $\{s_1, s_2\}$, with equal probabilities i.e. 0.5, then the entropy is $H(S) = 0.5 \log_2\left(\frac{1}{0.5}\right) + 0.5 \log_2\left(\frac{1}{0.5}\right) = 1$ which corresponds to a thorough *random* sequence. Our proposed system displays an average entropy value of 0.99358 which is high enough so that the system can resist the attacks and can possess an appropriate randomness.

Taking advantage of PRNGs to produce the keys, we made use of NIST SP 800-22 test to examine the randomness of the cipher [14]. Test Suit is a statistical package which consists of 16 tests. It was developed to test the randomness of binary sequences with arbitrary lengths produced by hardware/software-based cryptographic random or PRNGs.

Table IX. NIST test results

| Statistical test | | P-value | Results |
|---|---|---|---|
| Monobit | | 2 | Success |
| Block frequency(m=128) | | 0.7410 | Success |
| Runs | | 0.0758 | Success |
| Rank | | 0.0407 | Success |
| Nonoverlapping templates (M=1032,B=110101010) | | 0.2617 | Success |
| Overlapping templates (m=9,M=933,B=110101010) | | 1 | Success |
| Serial | P-value1 | 0.1234 | Success |
| | P-value2 | 0.2345 | Success |
| Cumulative sums | Forward | 0.0345 | Success |
| | Reverse | 0.0435 | Success |
| Random Excursion (state x) | X = -4 | 0.7147 | Success |
| | X = -3 | 0.2736 | Success |
| | X = -2 | 0.8530 | Success |
| | X = -1 | 0.6060 | Success |
| | X = 1 | 0.9371 | Success |
| | X = 2 | 0.4215 | Success |
| | X = 3 | 0.5970 | Success |
| | X = 4 | 0.8902 | Success |
| Random Excursion Variant (state x) | X = -9 | 0.3456 | Success |
| | X = -8 | 0.2476 | Success |
| | X = -7 | 0.3601 | Success |
| | X = -6 | 0.4138 | Success |
| | X = -5 | 0.3261 | Success |
| | X = -4 | 0.2123 | Success |
| | X = -3 | 0.1876 | Success |
| | X = -2 | 0.3408 | Success |
| | X = -1 | 0.4795 | Success |
| | X = 1 | 0.5557 | Success |
| | X = 2 | 0.8918 | Success |
| | X = 3 | 0.9161 | Success |
| | X = 4 | 0.7552 | Success |
| | X = 5 | 0.7237 | Success |
| | X = 6 | 0.2008 | Success |
| | X = 7 | 0.8702 | Success |
| | X = 8 | 0.9031 | Success |
| | X = 9 | 0.8864 | Success |
| Entropy | | 0.0219 | Success |

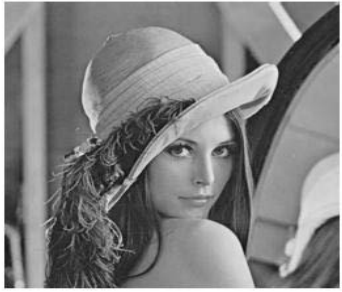 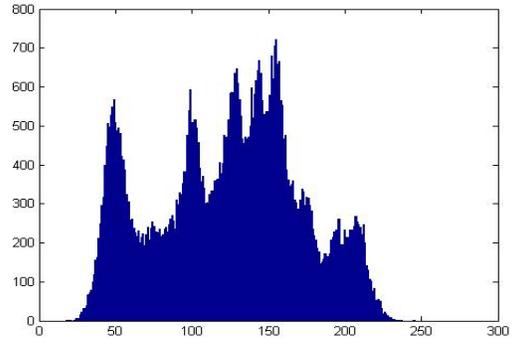

Figure4. Original Lena's image (256×256 pixel, 8-bit resolution) and its histogram before encryption

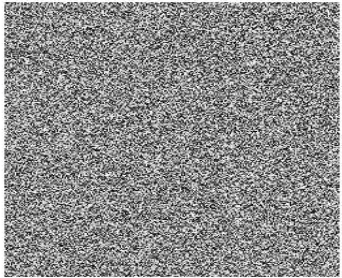 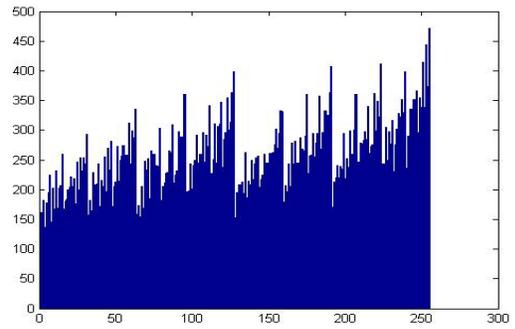

Figure5. Plain-image (256×256 pixel, 8-bit resolution) and the histogram after encryption

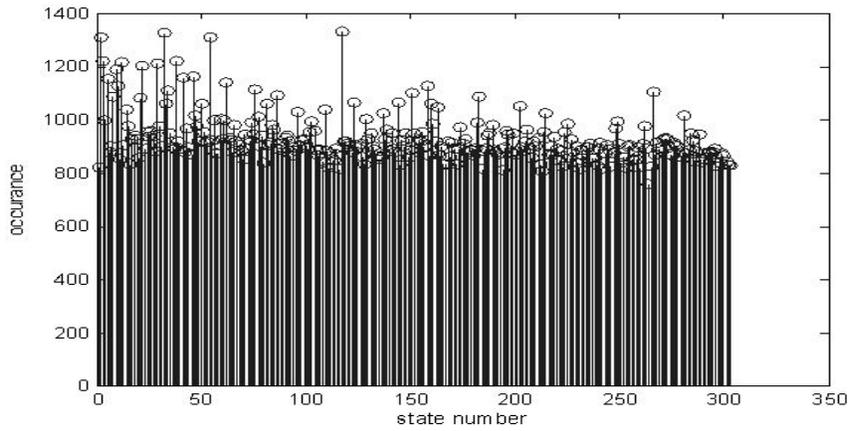

Figure6. Total times that each state have been met during encryption.

These tests focus on a variety of different types of non-randomness that could exist in a sequence. For instance, Table IX summarized the average of several results of testing several $10^6$-length cipher sequences. We can conclude from Table IX that the cipher which was encrypted from our proposed system is stochastic and it is robustness against known cipher-text attack.

### 4.3. Analysis of image encryption

A common way to study joint compression and encryption approaches characteristics is to apply it on images [15]. For image analysis, we have performed statistical analysis by calculating the histograms, the correlations of two adjacent pixels in the encrypted images and the correlation coefficient for several images and its corresponding encrypted images.

#### 4.3.1. Histogram

An image-histogram shows the distribution of different gray levels in an image. Fig.4 shows our plain image, Lena and its histogram. Clearly, histogram of a plain image contains large spikes which correspond to gray level values with most appearance. The histogram of the cipher image, shown in Fig.5 is more uniform and completely different from that of the original image. It is bears no statistical similarities to the plain image. Therefore the encrypted image histogram does not provide any clues to employ any statistical attack on the proposed image encryption procedure. In this experiment, we used a HFAC with N=7, P(0)=44/128 and Fmax=10 which produced 303 states, moreover we set 90% jump during transitions. Fig.6 clearly displays how many times each state has passed during encoding. The figure shows that all 303 states approximately are crossed 900 times. Consequently, it is not possible for a brute-force attacker to ignore some states to reduce the FSM. Since all 303 states are almost equally involved in encoding procedure

#### 4.3.2. Correlation coefficient between pixels

Calculating the correlation between adjacent pixels in an image is another way to measure the cipher image randomness. We calculated correlation between two vertically adjacent pixels, two horizontally adjacent pixels and two diagonally adjacent pixels in plain image and cipher image, respectively. First, we randomly selected 1000 pairs of two adjacent pixels from an image [16]. Then, we calculated their correlation coefficients using the following two formulas Where x and y are the values of two adjacent pixels in the image Eq.5 and Eq.6 [17]:

$$COV(x,y)=E\big((x-E(x))(y-E(y))\big), \quad (5)$$

$$r_{xy}=\frac{COV(x,y)}{\sqrt{D(x)D(y)}}, \quad (6)$$

where x and y are the values of two adjacent pixels in the image. Here the following discrete formulas were used:

$$E(x) = \frac{1}{N}\sum_{i=1}^{N} x_i ,$$

$$D(x) = \frac{1}{N}\sum_{i=1}^{N}(x_i - E(x))^2 ,$$

$$cov(x,y) = \frac{1}{N}\sum_{i=1}^{N}\big((x_i-E(x))(y_i-E(y))\big),$$

Fig.7 and Fig8. show the correlation distribution of two horizontally and vertically adjacent pixels in plain image and cipher image for the proposed coder respectively. The correlation coefficients between plain image and cipher image in horizontal, diagonal and vertical directions are shown in table X. It is clear that the two adjacent pixels in the plain image are highly correlated; however there is insignificant correlation between the two adjacent pixels in the cipher image.

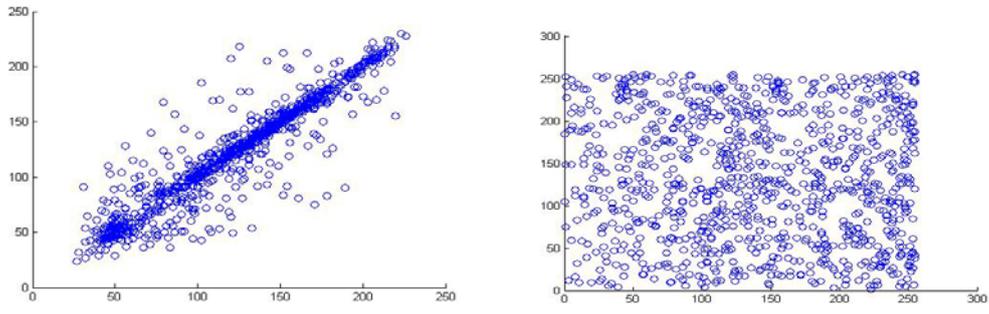

Figure7. Horizontal correlation between every two adjacent pixels (x,y) and (x+1,y) in the original left image and the right cipher image

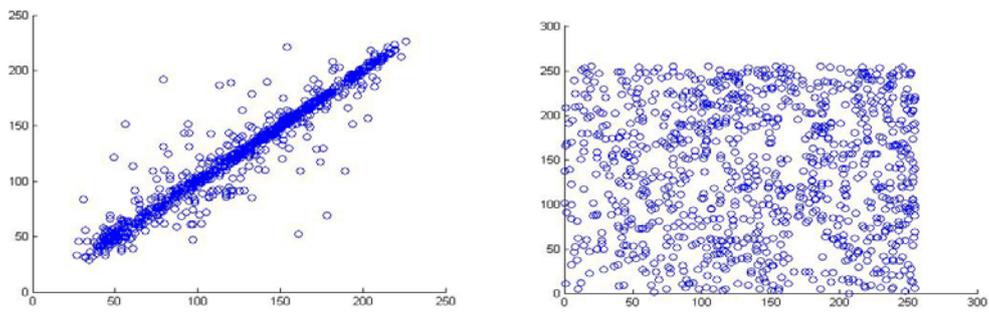

Figure8. Vertical correlation between every two adjacent pixels (x,y) and (x,y+1) in the original left image and the right cipher image

Table X. Correlation coefficients between Plain image and cipher image

| Direction of adjacent pixels | plaintext | ciphertext |
|---|---|---|
| Horizontal | 0.9856 | 0.0431 |
| Vertical | 0.9925 | 0.0315 |
| Diagonal | 0.9877 | -0.0014 |

Table XI. Correlation coefficients between encoded images with a slight change in keys

| Keys with one bit changed in them | | correlation coefficients |
|---|---|---|
| 1st image | 2nd image | |
| K1 | No bits | 1.5259 e-05 |
| K2 | No bits | 0.00002 |
| K1 | K2 | -0.0003 |
| K1 and K2 | No bits | 1.5259 e-05 |

## 4. 4. Sensitivity analysis

An ideal image encryption should be sensitive with respect to both the secret key and plain image. The change of a single bit in either the secret key or plain image should produce a completely different encrypted image. In the upcoming section, the finding obtained from the analysis of image encryption is reported.

### 4.4.1. Key sensitivity analysis

High key sensitivity in secure image cryptosystems guarantees that the cipher image cannot accurately be decrypted although there is only a slight difference between encryption or decryption keys. In addition it is an obstacle for brute-force attacks to some extent. In this study, the swap key was inserted to improve statistical characteristics of output sequence. It has not an overall effect on the encrypted sequence and it just swaps the output of each state. Consequently, we examine the sensitivity of encoder to changes in jump key. We performed the following steps:

- Changing one bit of *K1* which determine jump or not
- Changing one bit of *K2* which determines the next state
- Changing only one bit of the above two keys *K1* and *K2*.

It is not easy to compare the encrypted outputs by simply observing them. Thus, for the comparison, we calculated the correlation [18] between the corresponding bits of the four encrypted data by Eq.7

$$C_r = \frac{N\sum_{j=1}^{N}(x_j y_j) - \sum_{j=1}^{N}x_j \cdot \sum_{j=1}^{N}y_j}{\sqrt{\left(N\sum_{j=1}^{N}x_j^2 - \left(\sum_{j=1}^{N}x_j\right)^2\right) \cdot \left(N\sum_{j=1}^{N}y_j^2 - \left(\sum_{j=1}^{N}y_j\right)^2\right)}} \quad (7)$$

where, $x_j$ and $y_j$ are the values of corresponding pixels in the encrypted image and original image.

We went through the above mentioned stages by examining several different keys. Then, we calculated the correlation coefficient for the encrypted image and original image by using Eq.7. In all cases, very small correlation coefficients between the corresponding outputs were obtained. For instance, the correlation coefficients between encoded images with K1 and K2 *keys* of jump key for the outputs from the steps (a) to (c) are shown in Table XI.

### 4.4.2. Sensitivity to plaintext

One of the distinctive features of the encryption system proposes in this study is that it is highly sensitive to the slightest change (i.e., a single change bit) in the plain image. To test the influence of one-pixel change on the plain image which are encrypted by the proposed coder, two common measures may be used: Number of Pixels Change Rate (NPCR) and Unified Average Changing Intensity (UACI). Let two ciphered images, whose corresponding plain image have only one pixel difference, be denoted by C1 and C2. We labeled the grayscale values of the pixels at pixel (i,j) in C1 and C2 by C1(i,j) and C2(i,j), respectively. A bipolar array, D, with the same size as images C1 and C2 is defined as follows:

$$D(i,j) = \begin{cases} 1 & ; \quad C1(i,j) = C2(i,j) \\ 0 & ; \quad C1(i,j) \neq C2(i,j) \end{cases}$$

The NPCR is defined as Eq.8:

$$NPCR = \frac{\sum_{i,j} D(i,j)}{W.H} \times 100\% \quad (3)$$

where W and H are the width and height of C1 or C2. The NPCR measures the percentage of different pixel numbers between plain image and cipher image. The UACI is defined as Eq.9

$$UACI = \frac{1}{W.H}\left[\sum_{i,j}\frac{C_1(i,j) - C_2(i,j)}{255}\right] \times 100\% \quad (9)$$

which measures the average intensity of differences between the two images. NPCR is %99.6490, showing thereby that the encryption scheme is very sensitive with respect to small changes in the plain image. UACI is %45.49 which is indicating that one bit modification yields a similar result to theory. It is desirable that

Changing one bit in the plaintext to make theoretically a 50% difference [18] in the bits of the cipher. For all these reasons, the proposed scheme of this study proves to be sensitive to the changes in the input, hence, an ideal coder. Therefore, attacks by analyzing the static property of cipher text are prevented in our scheme. Moreover, based on analysis of image encryption and sensitivity analysis, the proposed joint encoder is robust against chosen plaintext attacks.

A pseudorandom sequence is vulnerable to the known plaintext attacks; since there is a given known input sequence, the attacker can compare the joint source-channel coder and the proposed coded sequences and attempt to find the added subintervals and their locations. To increase the security, an efficient key distribution protocol could be also explored in our algorithm to provide a sufficient encryption.

## 5. Conclusion

In this article we proposed a new method to add encryption to FSAC. This method was based on jump to states dictated by a key which was only known to the transmitter and receiver. In order to cut the relations between the successive outputs, we changed the outputs of the transitions to Huffman. Due to the large number of states being chosen in this method, applying the brute force method was almost impossible in order to find the key. We have carried out statistical analysis, key sensitivity analysis and key space analysis to demonstrate the security of the new image encryption procedure. The proposed joint compression-encryption is so faster and less complicated in comparison to disjoint coders therefore we conclude with the remark that the proposed method is expected to be useful for real time image encryption and transmission applications. In future research the emphasis can be more on adding some extra redundancy to the code and subsequently using this redundancy in the decoder to correct errors. In addition more research on theoretical proof of security is necessary.

# References

[1] M. Grangetto, A. Grosso, and E. Magli, "Selective encryption of JPEG2000 images by means of randomized arithmetic coding," in Proc. IEEE 6th Workshop on Multimedia Signal Processing. Sie na, Italy, Sep. 2004, pp. 347–350.

[2] M. Grangetto, E. Magli, and G. Olmo, "Multimedia selective encryption by means of randomized arithmetic coding," *IEEE Trans . Multimedia*, vol. 8, no. 5, pp. 905–917, Oct. 2006.

[3] P.Teekaput,S. Chokchaitam, "Secure Embedded Error Detection Arithmetic Coding" in Proceedings of the Third International Conf. on Information Technology and Applications (ICITA'05),  pp. 568-571, 2005:

[4] J. Wen, H. Kim, and J. D. Villasenor, "Binary arithmetic coding with key-based interval splitting," *IEEE Signal Process. Lett.*, vol. 13, no. 2,pp. 69–72, Feb. 2006.

[5] H. Kim, J. Wen, J. Villasenor, "Secure Arithmetic Coding", IEEE TRANS. on Signal Processing, vol. 55, no. 5, pp. 2263-2272.May 2007.

[6] R. Bose, S. Pathak, "A Novel Compression and Encryption Scheme Using Variable Model Arithmetic Coding and Coupled Chaotic System", *IEEE Transactions on Circuits and Systems I*, vol. 53, no. 4, Apr. 2006. pp. 848-857.

[7] A. Moffat, R. M. Neal, and I. H. Witten, "Arithmetic coding revisited," ACM Transactions on Information Systems, vol. 16,no. 3, pp. 256–294, 1998.

[8] P. G. Howard and J. S. Vitter, "Practical implementations of arithmetic coding," Kluwer Academic Publishers, Image and Text Compression, vol. 13(7), pp. 85–112, 1992.

[9] P. G. Howard and J. S. Vitter. "Design and analysis of fast text compression based on quasi-arithmetic coding," Inform. Proc. and Management, Vol. 30, No.6, pp. 777-790, Jun. 1994.

[10] H. Moradmand, A. Payandeh, M.R. Aref, "Joint Source-Channel Coding using Finite State Integer Arithmetic Codes," Proc. of 2009 IEEE International Conference on Electro/Information Technology(eit'09),pp 19-22, Windsor, ON, Canada, June 2009.

[11] S. Ben-Jamaa, C. Weidmann, and M. Kieffer, "Asymptotic error-correcting performance of joint source-channel schemesbased on arithmetic coding," in Proceedings of IEEE 8th Workshop on Multimedia Signal Processing (MMSP '06), pp. 262–266, Victoria, BC, Canada, October 2006.

[12] Moradmand H., Payandeh A,"Secure finite state integer arithmetic codes ," Proc. of the 2011 International Conference on of Advanced Technologies for Communications (ATC), pp. 10-13, Da Nang, Vietnam, August 2011.

[13] M. Sinaie, V.T. vakili,"secure arithmetic coding with error detection capability",Eurasip journal on information security,Vol.2010,Article ID. 621521 ,9 pages,2010

[14] A. Rukhin, J. Soto, J. Nechvatal et al., "A statistical test suite for random and pseudorandom number generators for cryptographic applications," NIST Special Publication 800-22, May 2001.

[15] Jianyong Chen, Junwei Zhou, Kwok-Wo Wong," a modified chaos-based joint compression and encryption", IEEE TRANSACTIONS ON CIRCUITS AND SYSTEMS—II: EXPRESS BRIEFS, VOL. 58, NO. 2, FEBRUARY 2011, pp.110-114.

[16] Hengjian Li, Jiashu Zhang,, A secure and efficient entropy coding based on arithmetic coding",CommunNonlinear Sci Numer Simulat 14 (2009) 4304− 4318

[17] N.K. Pareek, Vinod Patidar, K.K. Sud," Image encryption using chaotic logistic map", Image and Vision Computing 24 (2006) 926–934

[18] X. Tong,M. Cui, and Z.Wang, "A new feedback image encryption scheme based on perturbation with dynamical compound chaotic sequence cipher generator," Optics Communications, vol. 282, no. 14, pp. 2722–2728, 2009.